\documentclass[12pt,a4paper]{article}

\usepackage[greek,american]{babel}
\usepackage{amssymb}
\usepackage{amsfonts}
\usepackage{subfigure}
\usepackage[dvips]{graphicx}

\begin{document}

\begin{center}
{{{\Large\sf\bf A High Accuracy Stochastic Estimation of a Nonlinear Deterministic Model}}}\\

\vspace{0.5cm}
{\large\sf Spyridon J. Hatjispyros$^*$, Stephen G. Walker$^{**}$ }\\
\vspace{0.2cm}
\end{center}
\centerline{\sf $^*$ University of the Aegean, Karlovassi, Samos, GR-832 00, Greece.}  \centerline{\sf $^{**}$ University of Kent,
Canterbury, Kent, CT2 7NZ, UK. }

\begin{abstract} In this paper, an approach to estimating a nonlinear deterministic model is presented. We introduce a stochastic model with extremely small variances so that the deterministic and stochastic models are essentially indistinguishable from each other. This point is explained in the paper. The estimation is then carried out using stochastic optimisation based on Markov chain Monte Carlo (MCMC) methods.

\vspace{0.1in} \noindent {\sl Keywords:}  Logistic model; Nonlinear model; MCMC; Stochastic optimisation.
\end{abstract}

\section{Introduction}

\subsection{Background}
Given a time series coming from a physical process of interest,
the researcher is concerned with inferring
unknown quantities, thus modeling the unknown dynamical system in 
such a way that both systems, the physical and the proposed one,
share the same qualitative features.

In recent years, it has been recognized that purely deterministic systems,  
even with a small number of degrees of freedom, can lead to
complicated and irregular behavior. For example, types of nonlinear--chaotic systems
have been used in modeling a variety of physical phenomena.
Cases include: a simple model for turbulence (Ruelle and Takens, 1971),  
understanding the behavior of plasmas (Lichtenberg and Lieberman,
1983), the formation of prices in the stock market (Day, 1994 and
1997). These models can even incorporate dynamic noise as model error 
resulting in what is known as random dynamical systems (Lasota and Mackey,
1994).

In dynamical systems theory, when a functional form for the model has
been chosen, accurate estimation of quantities such as
the control parameters and initial conditions is crucial.
For example, when stochastic components are present, the functional
form of the deterministic part is polynomial and the information for 
the system is given in the form of a time series, a number of methodologies
using Bayesian, parametric or nonparametric, probability models have been proposed 
(Meyer and Christensen, 2000 and 2001, Hatjispyros et al. 2007 and 2009).

In this paper, we develop the methodology for obtaining high accuracy 
estimates of the control parameter and the initial condition when the proposed
system is a discrete time, one--dimensional nonlinear deterministic dynamical system.
The time series information is given in the form of a highly censored
sequence of observations, or equivalently a kneading sequence 
(Metropolis et al. 1973, Wang and Kazarinoff, 1987). 

In Wu et al. (2004), an alternative algorithm is presented for the estimation 
of the control parameter and the initial condition when a two--valued symbolic
sequence generated by iterations of the quadratic map is given. This algorithm utilizes
the relationship between the Gray Ordering Numbers and one--dimensional quadratic
maps introduced in Alvarez et al. (1998).

Due to the use of MCMC methods, which may be not familiar to the readers, we provide a brief
explanation here.

\subsection{MCMC and the Gibbs sampler} The approach to estimation of the dynamical system will rely on stochastic models and the sampling of these models. These samples are then used for estimation purposes. Markov chain Monte Carlo methods (Hastings 1970; Smith and Roberts, 1993; Tierney, 1994) are now routinely used in statistical inference problems, notably for Bayesian inference, when the models are complicated and direct estimation via numerical methods is unfeasible.   

Suppose the density function $f_{X,Y}(x,y)$ is to be sampled and direct sampling is not possible. 
A Markov chain is set up with stationary density $f_{X,Y}(x,y)$. Hence, as the chain proceeds, the samples taken, say $(x^{(i)},y^{(i)})$, 
converge to samples from this density and given a large number of these samples, say $N_G$, estimates of desired functionals can be computed. 
For example, the marginal mean of $x$ can be computed as
$$
{1\over N_G}\sum_{i=1}^{N_G} x^{(i)},
$$
where $x^{(i)}$ is the output from the chain for the $x$ values.   

A simple chain with the correct stationary density can be constructed in the following way. 
Start with a fixed point $x^{(0)}$ and sequentially sample, for $i=1,2,\ldots$, from the conditional densities
$$
f_{Y|X}(y|x^{(i-1)}),
$$
to get $y^{(i)}$, and then 
$$
f_{X|Y}(x|y^{(i)}),
$$
to get $x^{(i)}$. See Smith and Roberts (1993) for the details of the theory for such a chain, known specifically as the Gibbs sampler. 

To remove the bias of the starting point $x^{(0)}$, the first $N_b$ samples can be ignored and so inference relies only on the output
$$
\{(x^{(i)},y^{(i)})\}_{N_b<i\leq N_G}.
$$

Of course this sampler relies on the availability of the two conditional densities to be sampled directly. Often this can be the case. When it is not then a recourse is to introduce a latent variable, $z$ say, (Damien et al., 1999) which does not change the marginal density of $f_{X,Y}(x,y)$. The joint $f_{X,Y,Z}(x,y,z)$ is introduced so that
$$
\int_{\mathbb R} f_{X,Y,Z}(x,y,z)\,dz=f_{X,Y}(x,y),
$$
and an augmented Gibbs sampler operates now on $(x,y,z)$. This works by sequentially sampling the conditional densities,
$f_{X|Y,Z}(x|y,z)$, $f_{Y|X,Z}(y|x,z)$ and $f_{Z|X,Y}(z|x,y)$. The $(x,y)$ output can still be used for inference purposes. 

\subsection{Layout of paper} The layout is as follows. In Section $2$ we provide the probability model for the observations which contain the latent variables. The joint probability model therefore contain the parameters to be estimated and the unobserved variables. In Section $2$ we also describe the full conditional densities required to implement the Gibbs sampler. Details of how to sample these and latent variables useful for the implementation are provided in the Appendix.

Section $3$ then elaborates on the MCMC method which necessarily needs to be non--standard. This is because the joint density 
we are actually sampling is highly modal and with sharp and highly peaked modes. In the end we are searching the largest mode 
and for estimation purposes only, so we can ``zoom" in on the correct mode. Therefore the actual algorithm involves a number 
of steps, which are described in Section $3$. Further elaboration is provided in Section $4$ where we undertake an illustrative 
example involving the quadratic map.     

\section{The probability model}

Suppose we have a one dimensional nonlinear dynamical model 
$g:X\to X$, whereby for some initial condition $y_0$, 
the trajectory $y_i^*$ is formed from $y_{i-1}^*$ via the deterministic recurrence relation
$$
y_i^*=g(\vartheta,y_{i-1}^*),\quad i=1,\ldots,n.
$$
The map $g$ is continuous in $y^*$,
$\vartheta\in\Theta$ is the parametrization of the map and $X$ is invariant under $g(\vartheta,\;\cdot)$
for all values of $\vartheta\in\Theta$. We choose $g$ to be quadratic in $y_{i-1}$. 

The underlying data $y^*=(y_i^*)_{1\le i\le n}$, are censored. So if $y_i^*\in A$ we observe $b_i=0$ 
whereas if $y_i^*\notin A$ then we observe $b_i=1$. Given the binary symbolic data sequence of $b=(b_i)_{1\le i\le n}$ our 
aim is to estimate $(\vartheta,y_0)$. 

To permit a stochastic approach to estimation we adopt a stochastic model for the unobserved sequence $y^*$. 
So let the stochastic counterpart of $y_i^*$ be written as $y_i$. Therefore, for some specified sequence of variances $(\sigma_i^2)$, we assume that
$$
y_i=g(\vartheta,y_{i-1})+\sigma_i\epsilon_i,
$$
where the $(\epsilon_i)$ are independent and identically distributed as standard normal random variables.
The stochastic model can be made arbitrarily close to the deterministic model by allowing the variances $\sigma_i^2$ 
to collapse to zero. In practice, due to the nature of the data $(b_i)$, we only need to choose the variances small 
enough to ensure that the stochastic model gives rise to the same data with very high probability.
 
Here we start to describe some of the joint densities of interest that we would wish to sample in order to estimate the model. 
These are intractable from a numerical point of view but we will resolve this by using the Gibbs sampler.  
The random variables $y_i$, given $(\vartheta,y_{i-1},\sigma_i)$ are conditionally independent  
normals with means $g(\vartheta,y_{i-1})$ and variances $\sigma_i^2$; our notation for this will be
$$
y_i|\,\vartheta,y_{i-1},\sigma_i\sim N\left(g(\vartheta,y_{i-1}),\sigma_i^2\right),
\quad i=1,\ldots,n,
$$ 
and the joint density of $(y_1,\ldots,y_n)$ conditional on the values $y_0$, $\vartheta$ and $(\sigma_i)$ is
\begin{equation}
\label{NormalModel}
\pi(y_1,\ldots,y_n|\,y_0,\vartheta,(\sigma_i))=\prod_{i=1}^n \pi(y_i|y_{i-1},\vartheta,\sigma_i)\propto \prod_{i=1}^n 
\exp\left\{-{1\over 2\sigma_i^2}\, (y_i-g\left(\vartheta,y_{i-1}\right))^2\right\}.
\end{equation}

We observe sets $D_i$ where $D_i=A$ or $D_i=A^c$. 
So $D_i=A$ if $b_i=0$ and $D_i=A^c$ if $b_i=1$.
The joint probability model for $\{(D_i,y_i)\}_{1\le i\le n}$ is given by
$$
\pi(y_1,\ldots,y_n,D_1,\ldots,D_n|\,y_0,\vartheta,(\sigma_i))\,=\,
\pi(y_1,\ldots,y_n|\,y_0,\vartheta,(\sigma_i))\prod_{i=1}^n {\rm Pr}(D_i|\,y_i).
$$
The conditional probability of $D_i$ given $y_i$ is $1$ whenever $y_i\in D_i$ and
$0$ elsewhere. We express that by using the characteristic function notation 
$$
{\rm Pr}(D_i|\,y_i)={\cal I}(y_i\in D_i).
$$
Then the joint probability model becomes
\begin{equation}
\label{JointModel}
\pi(y_1,\ldots,y_n,D_1,\ldots,D_n|\,y_0,\vartheta, (\sigma_i))\propto\,\prod_{i=1}^n {\cal I}(y_i\in D_i)
\exp\left\{-{1\over 2\sigma_i^2}\, \left(y_i-g\left(\vartheta,y_{i-1}\right)\right)^2\right\}.
\end{equation}

One way to make this model accurate is to take the 
specific sequence of standard deviation constant $(\sigma_i=\sigma)$ and to make $\sigma$ 
very small. This follows since the $(y_i)$ are deterministically generated, albeit 
with rounding errors, which can be used to justify the inclusion of some small $\sigma$.

However, in this case, the likelihood function is not only multi--modal, but incredibly spiked at the modes. 
This renders direct estimation of $(\vartheta,y_0)$ very difficult, if not impossible. 
This is essentially caused by the fact that the combination of:
\begin{description}
\item 1. The noninformative prior specification of $\vartheta$ and $y_0$ for 
model map $g$, namely
\begin{equation}
\label{WeakPriors}
\pi(\vartheta)={\cal U}(\vartheta|\Theta),\quad \pi(y_0)={\cal U}(y_0|X),\quad
\end{equation}
here formulated in terms of prior distributions, where ${\cal U}(\cdot|A)$ denotes the uniform distribution 
over an appropriate subset $A$ of $\mathbb R$.
\item 2. The available sample of length $n$ of censored observations of the form
%\begin{eqnarray}
%\label{DataSet}
%   D_i(\vartheta,y_0)\,=\,\left\{\begin{array}{ll}
%                  [0,\infty)  & b_i=1\\
%                  (-\infty,0)  & b_i=0        \\       
%                   \end{array}\right.,\quad i=1,\ldots,n,
%\end{eqnarray}
\begin{eqnarray}
\label{DataSet}
   D_i(\vartheta,y_0)\,=\,\left\{\begin{array}{ll}
                  A^c  & b_i=1\\
                  A  & b_i=0        \\       
                   \end{array}\right.,\quad i=1,\ldots,n,
\end{eqnarray}
\end{description}
makes inefficient a direct $(\vartheta,y_0|D)$--inference 
via Markov chain Monte Carlo (MCMC) methods. Therefore we need to think about smart MCMC methods.

\section{An augmented MCMC method}

We first give some definitions:

\vspace{0.1in}\noindent{\bf Definition $1$}.
{\sl 
A normal ($g,\theta,D$)--candidate $n$--vector $\bar{y}=(\bar{y}_0,\bar{y}_1,\ldots,\bar{y}_n)$ 
for a one dimensinal map $g$ given a value of the parameter 
$\theta\in\Theta$ and a symbolic data set 
$D=D(\vartheta,y_0)=(D_1,\ldots,D_n)$ of the form (\ref{DataSet}), is the mean 
of the conditional distribution $\pi(y_0, y_1, \ldots, y_n|\,\theta, \sigma, D)$ under the normal model 
(\ref{NormalModel}); that is
$$
\bar{y}_i\,=\,{\mathbb E}(y_i|\,\theta, \sigma, D)\,=\,
\int_{{\mathbb R}^{n+1}}y_i\,\pi(y_0, y_1, \ldots, y_n|\,\theta, \sigma, D)\,dy,\quad
i=0,1,\ldots,n.
$$
}

\vspace{0.1in}\noindent{\bf Definition $2$}.
{\sl 
The estimating strength (ES) of a ($g,\theta,D$)--candidate point $\bar{y}_i$ for $i=0,\ldots,n$ is the number 
\begin{eqnarray}
{\cal S}_i(\theta|D)\,=\,\left\{\begin{array}{ll}
     {\rm min}\left\{1\le k\le n-i:g^{(k)}(\theta,\bar{y}_i)\in D_{i+k},\;
                              g^{(k+1)}(\theta,\bar{y}_i)\in D_{i+k+1}^c\right\} & 0\le i\le n-2\nonumber\\
                              {\cal I}\left(g(\theta, \bar{y}_{n-1})\in D_n\right)      & i=n-1\nonumber\\
                              0                                            & i=n.\\
                                \end{array}\right. 
\end{eqnarray}
The cumulative estimating strength (CES) of a $n$--vector candidate $\bar{y}$ is the sum 
\begin{equation}
\label{cumstrength}
{\cal S}(\theta|D)=\sum_{i=0}^{n-1} {\cal S}_i(\theta|D).
\end{equation}
}

Our strategy will be to maximize the CES function ${\cal S}(\theta|D)$ over $\Theta$, 
via MCMC methods. 
In turn, the approximation of a global maximum, if it exists, of the CES function
using a sample of censored observations of length $n$, over a 
$\Theta$--grid of length $\delta$,  and the use of the inverse of the map $g$ over $D$,
will provide us with approximations 
$(\theta^{\,n,\delta}, \bar{y}_0^{\,n,\delta})$ to the true values $(\vartheta, y_0)$ responsible  
for the sequence of censored observations $D$. In other words, we will be able to 
specify numerically a subset $\Theta^{n,\delta}$ of the parameter space, containing the true 
value $\vartheta$, along with an approximation to the 
($g,\theta^{\,n,\delta},D$)--candidate vector $\bar{y}^{\,n,\delta}$.
The latter will enable us to substitute the prior information for $\theta$ with 
$\pi(\theta)={\cal U}(\theta|\Theta^{\,n,\delta})$ and the initial data sets $(D_i)$ with 
the refined sets
$$
D_i^{\,n,\delta,\epsilon}=(\bar{y}_i^{\,n,\delta}-\epsilon,\bar{y}_i^{\,n,\delta}+\epsilon)\cap D_i,
$$
for appropriate small $\epsilon$, thus obtaining an efficient
$\theta$--augmented MCMC scheme that improves the accuracy
of the previous estimates $(\theta^{\,n,\delta},\bar{y}_0^{\,n,\delta})$.

\subsection{The strength maximizing MCMC}
We now describe the MCMC algorithm for the estimation of the truncation set $\Theta^{n,\delta}$ and the candidate
approximation $\tilde{y}^{n,\delta}$.

Using (\ref{JointModel}) with $\sigma_i=\sigma$ the conditional distribution 
$\pi(y_0, y_1, \ldots, y_n|\,\theta, \sigma, D)$ becomes
$$
\pi(y_0, y_1, \ldots, y_n|\,\theta, \sigma, D)\,\propto\,
\pi(y_0)\;\prod_{i=1}^n {\cal I}(y_i\in D_i)
\exp\left\{-{1\over 2\sigma^2}\, (y_i-g\left(\theta,y_{i-1}\right))^2\right\}.
$$
where the prior distribution for $y_0$ is uniform in the invariant set $X$ of $g$ as in (\ref{WeakPriors}). 
The Gibbs sampler works by sampling from the full conditional distributions in turn. Hence, we need to 
describe and sample iteratively from the densities:
\begin{description}
\item 1. The full conditional for $y_0$ is
$$
\pi(y_0|\cdots)\,\propto\, {\cal I}(y_0\in X)
\exp\left\{-{1\over 2\sigma^2}\, (y_1-g\left(\theta,y_{0}\right))^2\right\}.
$$
\item 2. The full conditionals for $y_i$, $1\leq i\leq n-1$ are 
$$
\pi(y_i|\cdots)\,\propto\, {\cal I}(y_i\in D_i)
\exp\left\{-{1\over 2\sigma^2}\, \left[ (y_i-g(\theta,y_{i-1}))^2
+(y_{i+1}-g(\theta,y_{i}))^2\right]\right\}.
$$
\item 3. The full conditional for $y_n$ is
$$
\pi(y_n|\cdots)\,\propto\, {\cal I}(y_n\in D_n)
\exp\left\{-{1\over 2\sigma^2}\, (y_n-g\left(\theta,y_{n-1}\right))^2\right\}.
$$
\end{description}

The full conditional distributions for $y_0,\ldots,y_{n-1}$ are truncated versions of nonstandard
densities. The full conditional for $y_n$ is a $D_n$--truncated normal distribution with mean 
$g(\theta,y_{n-1})$ and variance $\sigma^2$. In all cases to
sample, we use a slice sampling technique. More details on the particular sampling scheme 
are given in the appendix.

\subsection{Polishing the MCMC}
Having computed the truncating set $\Theta^{\,n,\delta}$, and, the
more informative, data sets 
$D_i^{n,\delta,\epsilon}$ for $i=0,1,\ldots,n$ we describe the MCMC scheme for polishing the $(\theta^{\,n,\delta},\bar{y}_0^{\,n,\delta})$ approximations to higher accuracy.

We would like to sample from the posterior 
$$
\pi\left(\theta,y_0, y_1, \ldots, y_n|\, \sigma, D^{\,n,\delta,\epsilon}\right),
$$ 
for very small values of $\sigma$. 
Letting $h(y_{i-1},y_i;\theta)=(y_i-g(\theta,y_{i-1}))^2$ and $\lambda=(2\sigma^2)^{-1}$, we have
\begin{eqnarray}
\nonumber
 \pi\left(\theta, y_0, y_1, \ldots, y_n|\, \sigma, D^{n,\delta,\epsilon}\right) & \propto &
{\cal I}\left(\theta\in \Theta^{n,\delta}\right)\,{\cal I}\left(y_0\in D_0^{n,\delta,\epsilon}\right)\\
\nonumber
    &\times   &\prod_{i=1}^n {\cal I}(y_i\in D_i^{n,\delta,\epsilon})\; e^{-\lambda\, h(y_{i-1},y_i;\,\theta)},
\end{eqnarray}
so we need to describe and sample iteratively from the densities:

\begin{description}
\item 1. The posterior full conditional for $\theta$ is
\begin{equation}
\label{ThetaFullConditionals}
\pi(\theta|\cdots)\,\propto\, {\cal I}(\theta\in \Theta^{n,\delta})\;
\exp\left\{-\lambda\, \sum_{j=1}^n h(y_{j-1},y_j;\theta)\right\}.
\end{equation}
\item 2. The full conditionals for $y_i$, $0\leq i\leq n$, are
\begin{eqnarray}
\label{yFullConditionals}
\pi(y_i|\cdots)\,\propto\, {\cal I}\left(y_i\in D_i^{n,\delta,\epsilon}\right)
   \left\{\begin{array}{lll}
             \exp\{-\lambda\, h_\theta(y_0,y_1;\theta)\}                                & \,i=0  \\
             \exp\{-\lambda\,(h_\theta(y_{i-1},y_i;\theta)+h_\theta(y_{i},y_{i+1};\theta))\}   & 1\le i\le n\\  
             \exp\{-\lambda\, h_\theta(y_{n-1},y_n;\theta)\}                            & \,i=n.  \\     
          \end{array}\right.
\end{eqnarray}
\end{description}
It is not difficult to verify that when $g(\theta,y)$ is affine in $\theta$, 
say $g(\theta,y)=\alpha(y)+\beta(y)\,\theta$, the full 
conditional for $\theta$ is a normal truncated on the interval $\Theta^{n,\delta}$, with mean
$$
 \mu_\theta\,=\,{\sum_{j=1}^n(y_j-\alpha(y_{j-1}))\beta(y_{j-1})\over \sum_{j=1}^n\beta(y_{j-1})^2},
$$
and variance
$$
 \sigma_\theta^2\,=\,{\sigma^2\over \sum_{j=1}^n\beta(y_{j-1})^2}.
$$

In the next section we will provide an application of the augmented MCMC algorithm 
for the quadratic map.

\section{Illustrative example}
Our model map $g$ will be 
\begin{equation}
\label{QuadraticMap}
g(\theta,y)=1+\theta\,y^2,
\end{equation} 
with parameter space $\Theta=[-2,0)$, and invariant set $X=(-1,1)$. 
With censoring set $A=(-\infty,0)$ and because $A\cap X=(-1,0)$ the 
data set $D$ is simplified to
\begin{eqnarray}
\label{QuadraticDataSet}
   D_i\,=\,\left\{\begin{array}{ll}
                 [0,1)  & b_i=1\\
                  (-1,0)  & b_i=0.        \\       
                  \end{array}\right.
\end{eqnarray}
We simulate a symbolic sequence $D=D(\vartheta,y_0)$ of length $n=1000$ from (\ref{QuadraticMap}) using 
as a control parameter $\vartheta=-1.71$ and initial condition $y_0=0.8$. These 
are the quantities we wish to estimate.

Noninformative prior specifications for $\theta$  and $y_0$ are given by the densities
\begin{equation}
\label{PriorSpecifications}
\pi(\theta)={\cal U}(\theta|-2,0),\quad \pi(y_0)={\cal U}(y_0|-1,1).
\end{equation}

\subsection{Grid maximization results}
We begin by maximizing the CES function ${\cal S}(\theta|D)$. We
set $\sigma=10^{-3}$, this value of $\sigma$ is to balance the noninformative
prior specifications in (\ref{PriorSpecifications}) and the small ranges of the 
posterior marginal distributions for $(y_i)$ as smaller $\sigma$'s will cause the
corresponding MCMC to converge extremely slow. At this point we do not have to use
the full length sample, a smaller sample $D'$ of length $m=600$ is enough to estimate both
$\theta$ and $y_0$ to $4$ decimal places and to obtain a $\Theta^{m, \delta}$ truncating set of 
length $\delta\approx 3\times 10^{-3}$.

In our simulated example we consider the $\Theta$--grid sequence
in such way that the true value $\vartheta$ is not a grid point at any level.
Initially we consider the grid 
\begin{equation}
\label{ThetaGrid1}
\Theta_1=\{\theta_{0,1}=-1.500,\,\theta_{s,1}=\theta_{0,1}-0.045\, s,\,\,1\le s\le 11\},
\end{equation}
then for each value $\theta\in\Theta_1$ we compute iteratively via the strength 
maximizing MCMC scheme the vector of running averages 
$\bar{y}_{(j)}$ and the associated CES function ${\cal S}_{(j)}(\theta_r|\,D')$.
After a burn--in period of about $N_b=4\times 10^4$ iterations and $N_G=2\times 10^5$
in total Gibbs iterations we compute an approximation to the CES function as
$$
{\cal S}_{(N_G)}(\theta|\,D') \,=
\max_{N_b\,<\, j\,\le N_G}{\cal S}_{(j)}(\theta|\,D'),\quad\theta\in\Theta_1.
$$
The result is shown in Fig. $1$(a). The CES function has a maximum over $\Theta_1$
at $\theta_{6,1}=-1.7250$ which is the center of the interval 
$(-1.77,-1.68)$. The associated value for the estimate of the initial condition is 
$\bar{y}_0=0.80256$. This gives us the next refinement grid
\begin{equation}
\label{ThetaGrid2}
\Theta_2=\{\theta_{0,2}=-1.6800,\,\theta_{s,2}=\theta_{0,2}-0.0082\, s,\,\,1\le s\le 11\}.
\end{equation}
Reiterating the whole process, we find a maximum of the ${\cal S}_{(N_G)}(\theta|\,D')$ 
function over $\Theta_2$ at $\theta_{5,2}=-1.71273$ and $\bar{y}_0=0.80073$, as shown in Fig. $1$(b), 
which is the center of the interval $(-1.72091,-1.70455)$. Our final refinement with the
$D'$ data set is
\begin{equation}
\label{ThetaGrid3}
\Theta_3=\{\theta_{0,3}=-1.70455,\,\theta_{s,3}=\theta_{0,3}-0.00270\, s,\,\,s=1,\ldots, 11\},
\end{equation}
where we find a maximum at $\theta_{5,3}=-1.70995$ and $\bar{y}_0=0.80026$, 
see Fig. $1$(c). Both approximations agree to the true values
to $4$ decimal places (see table $1$). As a $\theta$--prior information to the polishing MCMC we 
are using the interval $(\theta_{6,3},\theta_{4,3})=(-1.71132,-1.70859)$.
A plot of the sequence of strengths at the maximum $\theta_{5,3}$ for $D'$ can be found in 
Fig. $2$. The sequence is highly oscillatory and a number of maximums are descernible. The
worst estimated $\bar{y}_i$ values are on the tail of the sequence.

%$$
%{\cal S}_k(\theta|D),\quad k,\quad {\cal S}(\theta|D'),\quad
%|y_i-\bar{y}_i|,\quad |y_i-\tilde{y}_i|
%$$
%$$
%{\mathbb E}(y_0|D)\quad{\mathbb E}(\theta|D), \quad {\rm MCMC~ iterations}
%$$
%$$
%\theta, y_0
%$$

\subsection{Increasing the accuracy of the $\bar{y}_i$ estimates}

In the sequel we are making use of the full sample $D$ of size $n=1000$. We compute
the sequence of strengths at the closest maximum $\theta^*$ to the true value $\vartheta$
 -- in our example is $\theta_{5,3}$ -- and we pick $\bar{y}_\kappa$
as the candidate point that
is closer to the end of the sequence and has a strength value greater than $20$.
We discard estimates of order higher than $\kappa$, and we apply 
$g^{-1}|_D(\theta^*,\,\cdot)$, the inverse of $g$ over $D$ at $\theta^*$, to the remaining
candidate points. For example, for the quadratic map in (\ref{QuadraticMap}), we have
\begin{eqnarray}
\label{InverseMapSample}
   \tilde{y}_{\kappa-j}\,=\,\left\{\begin{array}{ll}
   \bar{y}_\kappa  & \,\,j=0\\
   {\rm sgn}(\bar{y}_{\kappa-j})\sqrt{(\theta^{*})^{-1}(\tilde{y}_{\kappa-j+1}-1)}  & \,\,1\le j\le\kappa, \\       
                   \end{array}\right.
\end{eqnarray}
and in our example $\kappa=984$.

We plot the sequence $({\cal S}_i(\theta^*|D))_{0\le i\le 80}$ of the first $80$ strengths, 
including the strength of $\bar{y}_0$, in Fig. $3$(a). 
The  absolute error of the estimates $\bar{y}_i$ at $\theta^*$ for $0\le i\le 80$, is in Fig. $3$(b).
Finally, in Fig. $3$(c), we plot the absolute error of the estimates 
$\tilde{y}_i$, obtained by (\ref{InverseMapSample}).
We can see that the order of the errors after the application of $g^{-1}|_D$ to the candidate
points $(\bar{y}_i)$ are considerably reduced from order $10^{-3}$ to order $10^{-5}$ (see table $2$).
The values in Figures $3$(a) and $3$(b) are negatively correlated with Pearson's correlation
$\gamma=-0.214$, Kendall's $\tau=-0.388$ and Spearman's correlation $\rho=-0.500$.

\subsection{Obtaining the high accuracy estimates of $(\theta,y_0)$}
We first use a $100$ length sub--sample of $D$ with $\epsilon=5\times 10^{-5}$. We iterate the augmented
MCMC using as an initial condition the vector $(\theta^*, \tilde{y}_0, \tilde{y}_1, \ldots, \tilde{y}_{100})$
and $\sigma=10^{-8}$.
The approximations to the mean of the $\theta$ and $y_0$ posteriors 
are shown in Fig. $4$(a) and $4$(b) respectively. We can see that after a large number of iterations 
are still oscillating in the intervals $\theta\in[-1.7100024, -1.7100013]$ and $y_0\in[0.7999952, 0.7999951]$ 
and are correct to $5$ decimal places.

Finally we increase the sample size to $\kappa=984$. The posterior ergodic averages for $\theta$ 
and $y_0$ are shown in Fig. $5$(a) and $5$(b) respectively. Here we approximate the true values 
to $7$ decimal places. The values are: $\theta= -1.70999996$ and $y_0= 0.79999995$.

\section{\bf Discussion}

We have described an algorithm for the estimation of the parameters of a one--dimensional quadratic 
nonlinear dynamical system. The stochastic model
provides us with an objective function to be maximized. However, the function is multimodal with sharp spikes and hence standard application of stochastic search procedures based on MCMC methods will fail. We therefore introduce a zooming or polishing process which allows us to target the largest mode.
Estimating maximum of functions using MCMC methods is by now a 
standard practice and dates back to the work of Geyer and Thompson (1992) 
and Douset et al. (2002) for a Bayesian setting. 
The maximum can be obtained to an arbitrarily level of accuracy. 

Our proposed algorithm besides its practical use on estimating $(\theta, x_0)$ from a set of 
incomplete measurements can be applied to a chaotic cryptography setting Baptista 1998, Alvarez et al 2003
and references therein. 
In such systems $(\theta, x_0)$ is used as
a shared secret key, from both the encoder of the plain text and the decoder of the cipher text.
Reusing the same secret key to encrypt messages weakens the security of the cryptosystem.
We can think of a chaotic cryptosystem transmitting together with the coded message a long $m$-valued
symbolic sequence which is the projection of a $(\theta, x_0)$--orbit on an uneven $m$--partition of the 
invariant set of the map, being a refinement of some initial generating partition. The new partition 
then becomes a shared secret key for the recovery of the control parameters and initial condition
from the decoder.

Our method in principle can be applied for estimating more complex nonlinear recurrences.
Although the complexity of the algorithm will increase, the associated computational workload 
will not increase dramatically, for example:

\begin{enumerate}

\item 
When $g(\theta, y)$ has degree $m>2$ in $y$. Then there are $m-1$ critical points in the 
invariant set $X$ of $g$ and each
data set $D_i$ will be one of the $m$ sets of the partition of $X$ induced by the critical points 
of the map. Such a partition is generating, so that there is a one to one correspondence between 
$m$--valued symbolic sequences and dynamical orbits. The only fact that adds to the complexity of
our algorithm is that sampling from the full conditional distributions 
for $y_i$ in (\ref{yFullConditionals}) 
will now involve solving numerically inequalities of degree $m$ of the form
$$
y_{i+1}-\sqrt{u_{i2}}<g(\theta, y_i)<y_{i+1}+\sqrt{u_{i2}},
$$
where $u_{i2}$ is an auxiliary random variable, see appendix equation (\ref{uinequality}). 

\item 
Higher dimensional dynamical systems of the form
$y_i=g(\theta, y_{i-1},\ldots, y_{i-d})$. In this case there is no systematic way for the construction 
of a generating partition Eisele (1999). Nevertheless it is possible to design a partition that 
maximizes the entropy content of the associated symbolic sequences Strelioff and Cruchfield (2007), 
Ebeling and Nicolis (1991). Then the conditional distributions for $y_i$ for $i=d,\ldots,n-d$ are given by
$$
\pi(y_i|\cdots)\,\propto\, {\cal I}(y_i\in D_i)\prod_{j=0}^d
\pi(y_{i+j}|\,y_{i+j-1},\ldots,y_{i+j-d},\theta).
$$
Sampling from such full conditionals will involve embedded MCMC schemes with $d+1$ auxiliary random variables.

\end{enumerate}

\section*{Appendix}

Here we show how to sample from the posterior full conditional distribution
for $y_i$ for $0<i<n$ in (\ref{yFullConditionals}). The Gibbs samplers for the full conditionals of
$\theta, y_0$ and $y_n$ are all special cases of the sampler that follows.

We introduce positive auxiliary random variables $u_{i1}$ and $u_{i2}$ and define the augmented
conditional density
$$
\pi(y_i,u_{i1},u_{i2} |\cdots)\,\propto\, {\cal I}(y_i\in D_i^{n,\delta,\epsilon})\;
e^{-\lambda u_{i1}}{\cal I}(u_{i1} >  h_\theta(y_{i-1},y_i))\;
e^{-\lambda u_{i2}}{\cal I}(u_{i2} >  h_\theta(y_{i},y_{i+1})).
$$
It is clear that the $y_i$ marginal of the latter density is the density we want to sample from.
We now describe the associated Gibbs sampler.

The full conditional densities for the latent variables $u_{i1}$ and $u_{i2}$ are
%$$
%\pi(u_{i1} |y_i,u_{i2}\cdots)\,\propto\, 
%e^{-\lambda u_{i1}}{\cal I}(u_{i1} >  h_\theta(y_{i-1},y_i)).
%$$
$$
\pi(u_{ij} |\cdots)\,\propto\, 
e^{-\lambda u_{ij}}{\cal I}(u_{ij} >  h_\theta(y_{i+j-2},y_{i+j-1})),\quad j=1,2.
$$
Hence, $u_{i1}$ and $u_{i2}$ are both exponential variables, with mean $1/\lambda$, truncated in the intervals
$(h_\theta(y_{i+j-2},y_{i+j-1}),\infty)$ for $j=1,2$, and thus they can be sampled as
$$
u_{ij}\,=\,-{1\over\lambda}\log (q_{ij})+h_\theta(y_{i+j-2},y_{i+j-1}),
$$
where $q_{ij}\sim{\cal U} (0,1)$ independently.

The full conditional for $y_i$ is
$$
\pi(y_i |u_{i1},u_{i2},\cdots)\,\propto\, {\cal I}(y_i\in D_i^{n,\delta,\epsilon})\;
{\cal I}(u_{i1} >  h_\theta(y_{i-1},y_i))\;{\cal I}(u_{i2} >  h_\theta(y_{i},y_{i+1})),
$$
which is in general a mixture of uniforms. For example when $g(\theta,y)$ is a polynomial of degree
$m$, in $y$, the full conditional for $y_i$ is a mixture of at most $m$ uniforms. In the special case
when $g(\theta,y)=1+\theta y^2$ we have 
\begin{eqnarray}
\label{uinequality}
{\cal I}(u_{i2} >  h_\theta(y_{i},y_{i+1}))\,=\,
   \left\{\begin{array}{ll}
             {\cal I}(y_i\in B_i)                         & s_i\le 0  \\
             {\cal I}(y_i\in B_i^-)+{\cal I}(y_i\in B_i^+) & s_i > 0   \\      
          \end{array}\right.,
\end{eqnarray}
where $B_i^-=(-\sqrt{r_i},-\sqrt{s_i})$, $B_i^+=(\sqrt{s_i},\sqrt{r_i})$ and $B_i=(-\sqrt{r_i},\sqrt{r_i})$ 
with \\ $s_i=\theta^{-1}(y_{i+1}+\sqrt{u_{i2}}-1)$ and $r_i=\theta^{-1}(y_{i+1}-\sqrt{u_{i2}}-1)$. By letting
$$
C_i\,=\,D_i^{n,\delta,\epsilon}\cap
\left(g(\theta,y_{i-1})-\sqrt{u_{i1}},\,g(\theta,y_{i-1})+\sqrt{u_{i1}}\right), 
$$
we finally obtain the uniform mixture
\begin{eqnarray}
\pi(y_i |u_{i1},u_{i2},\cdots)\,=\,
   \left\{\begin{array}{ll}
             {\cal U}(y_i| C_i\cap B_i)                                        & s_i\le 0 \nonumber \\
             p\,{\cal U}(y_i| C_i\cap B_i^-)+(1-p){\cal U}(y_i| C_i\cap B_i^+) & s_i > 0            \\      
          \end{array}\right.,
\end{eqnarray}
with 
$$
p={|C_i\cap B_i^-|\over |C_i\cap B_i^-|+|C_i\cap B_i^+|}.
$$

\newpage

\vspace{0.2in} \noindent {\bf References.} 

\begin{description}

\item Alvarez G., Montoya F., Romera, M., Pastor G. (2003). Cryptanalysis of an ergodic chaotic cipher.
{\sl Physics Letters A} {\bf 311}, 172--9.

\item Alvarez G., Romera M., Pastor G., et al. (1998). Gray codes and 1D quadratic maps.
{\sl Electron Lett} {\bf 34}, 1304–-6.

\item Baptista, M.S. (1998). Cryptography with chaos.
{\sl Physics Letters A} {\bf 240}, 50--4.

\item Damien P., Wakefield, J., Walker, S.G. (1999).
Gibbs sampling for Bayesian non--conjugate and hierarchical models by using 
auxiliary variables. {\sl J. R. Statist. Soc. B} {\bf 61}, part {\bf 2}, 331--344.

\item Doucet A., Godsill S.J., Roberts C.P. (2002).
Marginal maximum a posteriori estimation using Markov chain Monte Carlo.
{\sl Statistics and Computing} {\bf 12}, 77--84

\item Ebeling W., Nicolis G., (1991).
Entropy of symbolic sequences: the role of correlations.
{\sl Europhys. Lett.} {\bf 14} (3), 191–-6.

\item Eisele M., (1999).
Comparisons of several partitions of the Henon map. 
{\sl J. Phys. A: Math. Gen.} {\bf 32}, 1533–-1545.

\item Geyer, C.J. and Thompson, E.A. (1992). Constrained Monte Carlo Maximum 
Likelihood for Dependent Data. {\sl  J. Roy. Statist. Soc. B} {\bf 54} 657--699.

\item Hastings, W.K. (1970). Monte Carlo sampling methods using
Markov chains with applications. {\sl Biometrika}  {\bf 57}, 97--109.

\item Hatjispyros, S.J., Nicoleris, T., Walker, S.G. (2007). 
Parameter estimation for random dynamical systems using slice sampling. 
{\sl Physica A} {\bf 381}, 71--81.

\item Hatjispyros, S.J., Nicoleris, T., Walker, S.G. (2009).
A Bayesian nonparametric study of a dynamic nonlinear model.
{\sl Journal of Computational Statistics and Data Analysis} {\bf 381}, 71--81.

\item Lasota, A., Mackey, M.C. (1994).
Chaos, fractals, and noise. {\sl Applied Mathematical Sciences} 
{\bf 97}. Springer--Verlag, New York.

\item Lichtenberg A., Lieberman M., Regular and Stochastic Motion, Springer, New York, 1983.

\item Madan R. (1993), Chua's circuit: a paradigm for chaos, {\sl World Scientific}.

\item Metropolis, N., Stein M., Stein P. (1973). 
On the limit sets for transformations on the unit interval
{\sl Journal of Combinatorial Theory (A)} {\bf 15}, 25–-44.

\item Meyer R., Christensen, N. (2000).
Bayesian reconstruction of chaotic dynamical systems. 
{\sl Phys. Rev. E}  {\bf 62}, 3535--3542.

\item Meyer R., Christensen, N. (2001).
Fast Bayesian reconstruction of chaotic dynamical systems via extended Kalman filtering 
{\sl Phys. Rev. E}  {\bf 65}, 016201--8.

\item  Ruelle D., Takens F. (1971). On the nature of turbulence, {\sl Comm. Math. Phys.} 
{\bf 20} (1971) 167-–192;
On the nature of turbulence, {\sl Comm. Math. Phys.} {\bf 23} 343-–344.

\item Smith A.F.M., Roberts G.O. (1993).
Bayesian computation via the Gibbs sampler and related Markov chain Monte Carlo methods
{\sl J. R. Statist. Soc. B} {\bf 55}, part {\bf 1}, 3--23.

\item Strelioff C.C., Cruchfield J.P. (2007).
Optimal instruments and models for noisy chaos
{\sl Chaos} {\bf 17}, 043127.

\item Wang, L., Kazarinoff, N.D. (1987). 
On the universal sequence generated by a class of unimodal functions
{\sl Journal of Combinatorial Theory, series A} {\bf 46}, 39–-49.

\item Wu X, Hu H, Zhang B (2004). 
Parameter estimation only from the symbolic sequences generated by chaos system
{\sl Chaos, Solitons and Fractals} {\bf 22}, 359--366.

\end{description}

%--------------------------------------------------------------------------------------------------
% STRENGTH
%--------------------------------------------------------------------------------------------------
\begin{figure}
\begin{center}
\mbox{
\subfigure[]{\includegraphics[angle=0,width=0.50\textwidth]{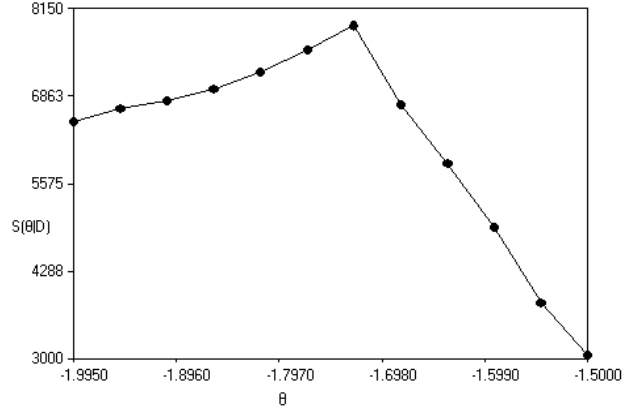}}}\\
\mbox{
\subfigure[]{\includegraphics[angle=0,width=0.50\textwidth]{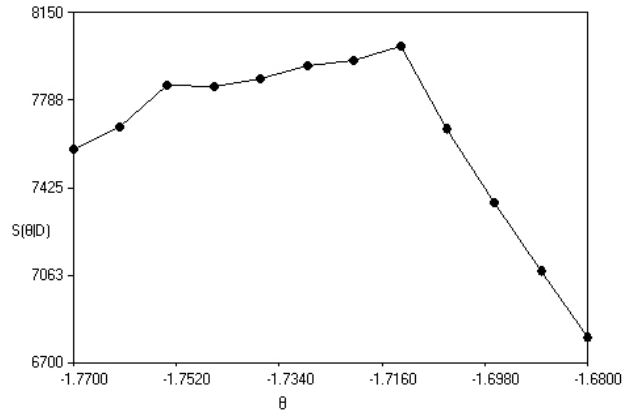}}}\\
\mbox{
\subfigure[]{\includegraphics[angle=0,width=0.50\textwidth]{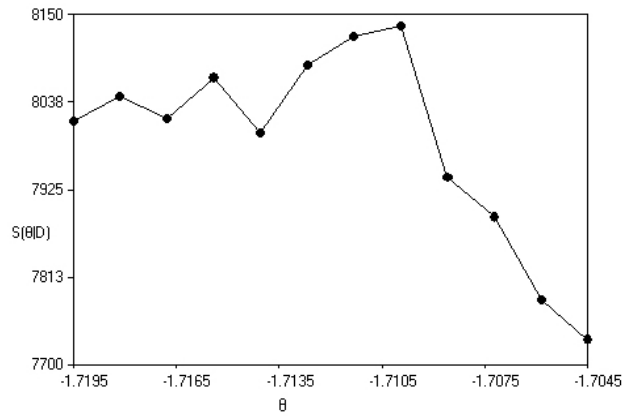}}}\\
\caption{Sequence of zooming process for mode seek. In subfigures (a)--(c) we maximize 
         the objective function ${\cal S}(\theta|D)$ in (\ref{cumstrength}) over the
         $\Theta$--grids in relations $(\ref{ThetaGrid1})$--$(\ref{ThetaGrid3})$. }
\label{fig:predictives}
\end{center}
\end{figure}

\begin{figure}
	\begin{center}
		\includegraphics[angle=0,width=0.75\textwidth]{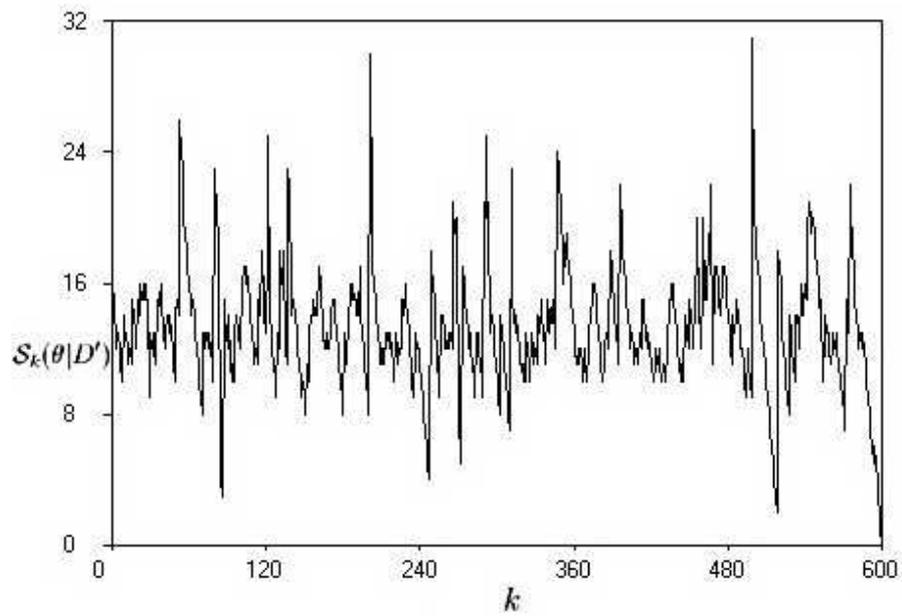}
	\end{center}
	\caption{The sequence of strengths  $\{{\cal S}_k(\theta|D')\}_{k=0}^{600}$ for $\theta=-1.70995$
	         and a simulated symbolic sequence $D'$ from the quadratic map (\ref{QuadraticMap}) of length 
	         $600$ with control parameter value at $\theta=-1.71$ and initial condition $y_0=0.8$. There are
	         $21$ strengths with value greater than $20$.}
	\label{fig:strenght600}
\end{figure}
%--------------------------------------------------------------------------------------------------
% ABSOLUTE ERRORS
%--------------------------------------------------------------------------------------------------
\begin{figure}
\begin{center}
\mbox{
\subfigure[]{\includegraphics[angle=0,width=0.5\textwidth]{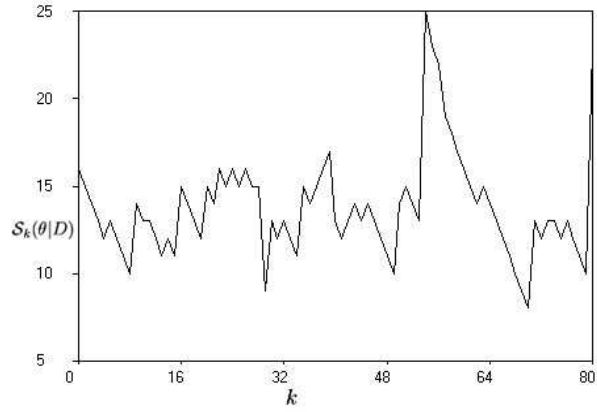}}}\\
\mbox{
\subfigure[]{\includegraphics[angle=0,width=0.5\textwidth]{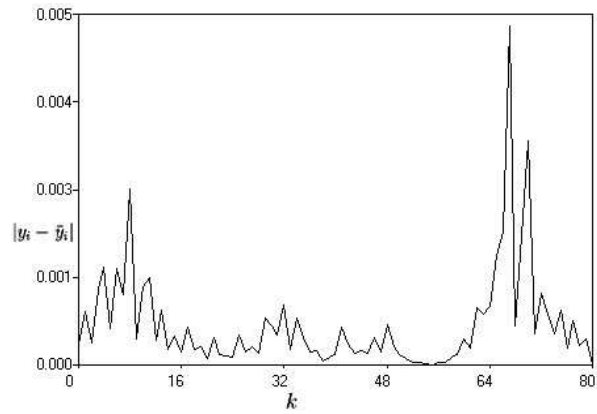}}}\\
\mbox{
\subfigure[]{\includegraphics[angle=0,width=0.5\textwidth]{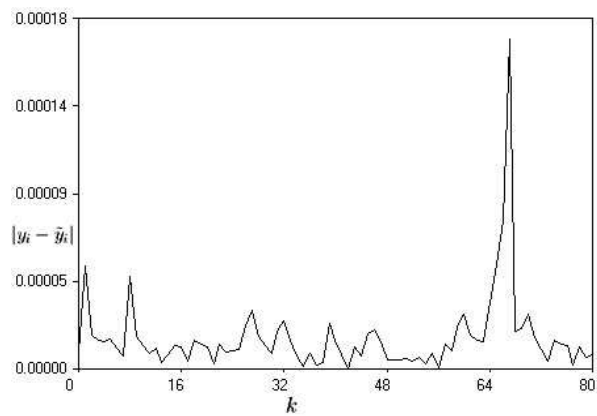}}}\\
\caption{In in Fig. $3$(a) we plot the sequence ${\cal S}_i(\theta^*|D)$ of individual strengths 
         for $0\le i\le 80$ based on a data set of length $600$. In Fig. $3$(b) we plot
         the absolute error of the estimates $\bar{y}_i$ at $\theta^*$. The sequence
         of strengths and absolute errors are negatively correlated. Finally in Fig. $3$(c)
         we plot the absolute error of the estimates $\tilde{y}_i$, obtained with the help of 
         (\ref{InverseMapSample}).}
\label{fig:predictives}
\end{center}
\end{figure}

%--------------------------------------------------------------------------------------------------
% 5 DECIMAL PLACES 
%--------------------------------------------------------------------------------------------------
\begin{figure}
\begin{center}
\mbox{
\subfigure[]{\includegraphics[angle=0,width=0.6\textwidth]{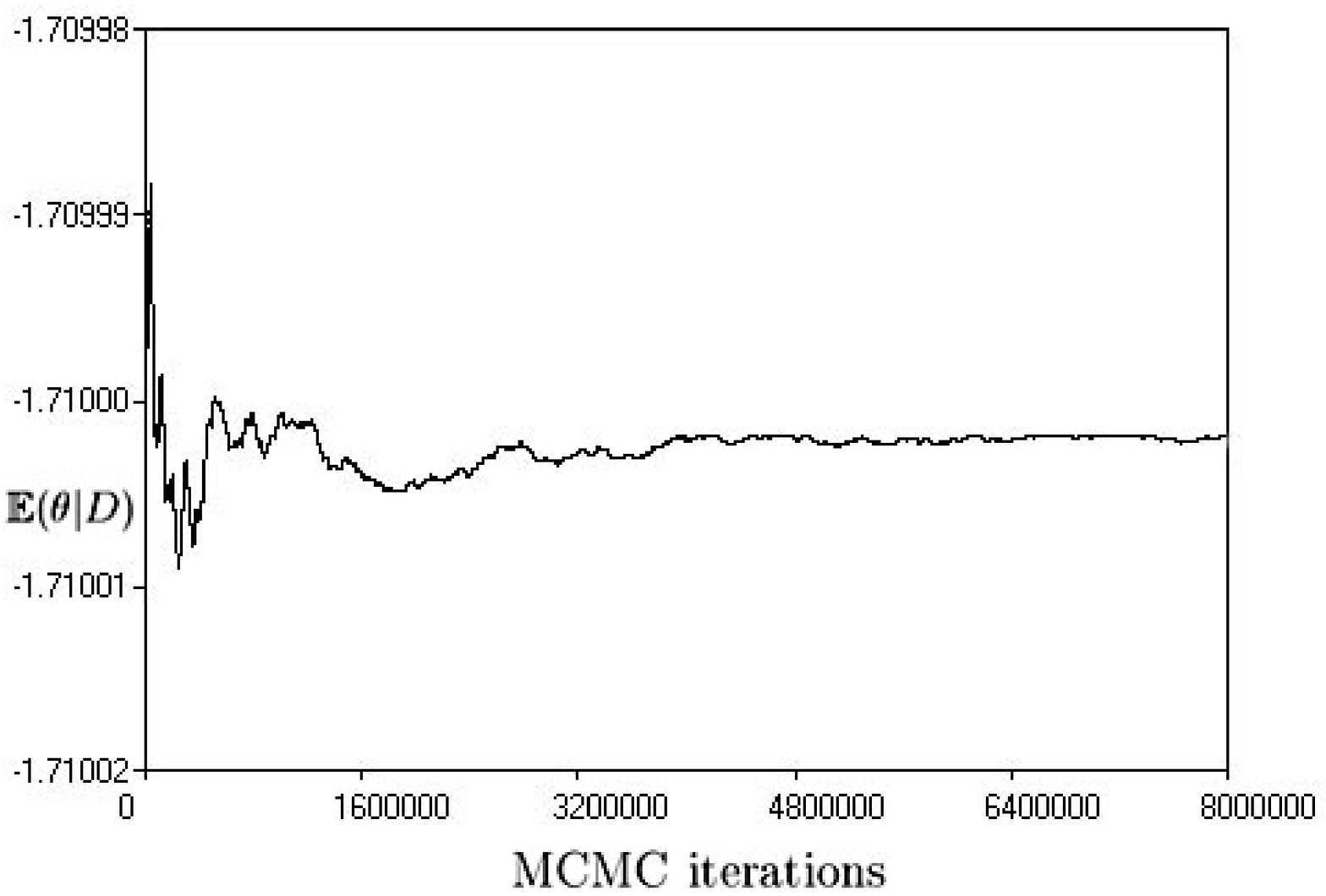}}}\\
\mbox{
\subfigure[]{\includegraphics[angle=0,width=0.6\textwidth]{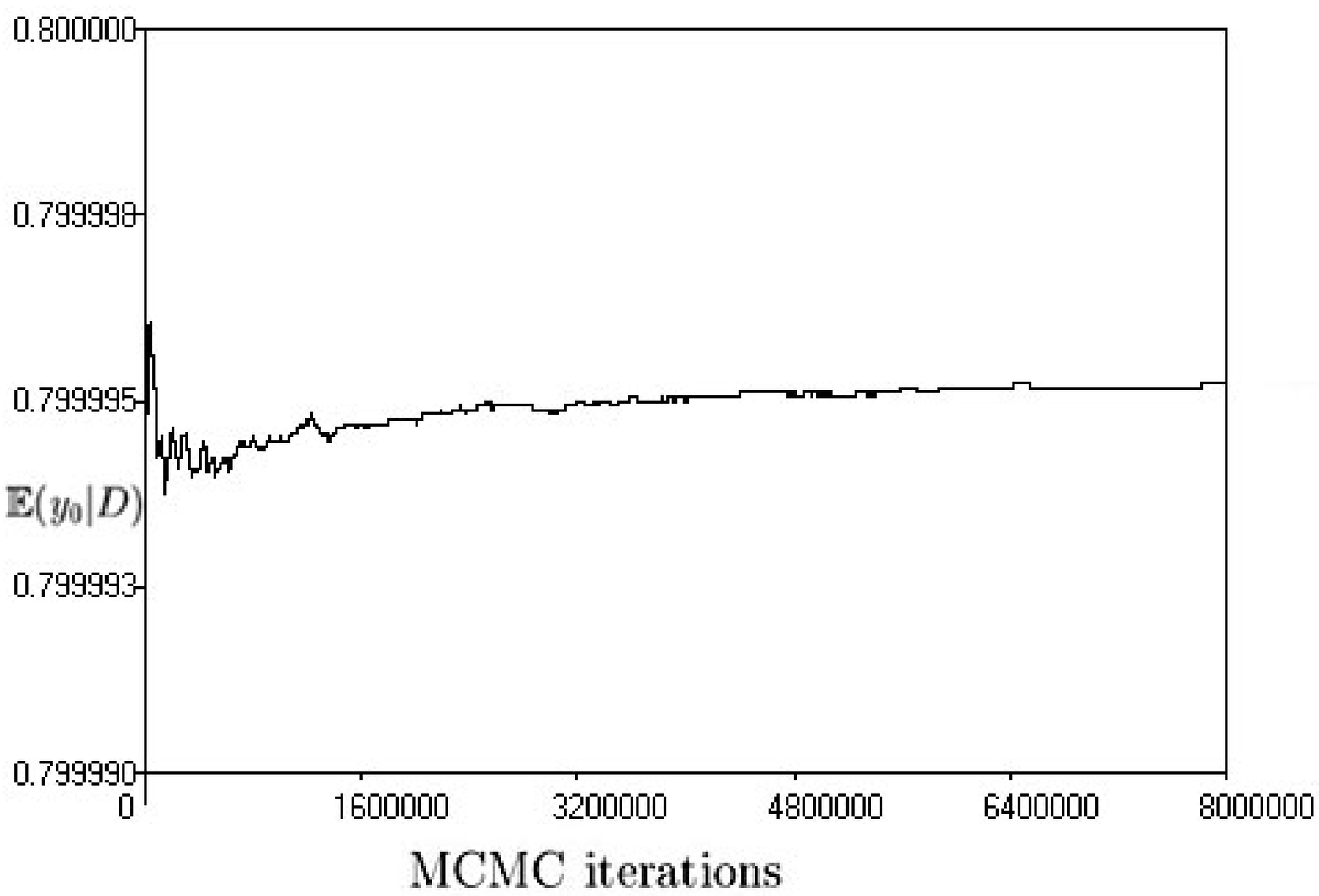}}}\\
\caption{Here the sample size is $100$ with $\epsilon=5\times 10^{-5}$ and $\sigma=10^{-8}$.
         The initial conditions for the MCMC scheme in 
         (\ref{ThetaFullConditionals})--(\ref{yFullConditionals}) is the vector 
         $(\theta^*, \tilde{y}_0, \tilde{y}_1, \ldots, \tilde{y}_{100})$ with $\tilde{y}_i$'s coming
         from  (\ref{InverseMapSample}).
         The approximations of the $\theta$ and $y_0$ posterior means
         -- shown in Fig. $4$(a) and $4$(b) respectively -- are converging to
         $\theta=-1.7100024$ and $y_0=0.7999952$ and are correct to $5$ decimal places.}
\label{fig:predictives}
\end{center}
\end{figure}

%--------------------------------------------------------------------------------------------------
% 7 DECIMAL PLACES 
%--------------------------------------------------------------------------------------------------
\begin{figure}
\begin{center}
\mbox{
\subfigure[]{\includegraphics[angle=0,width=0.6\textwidth]{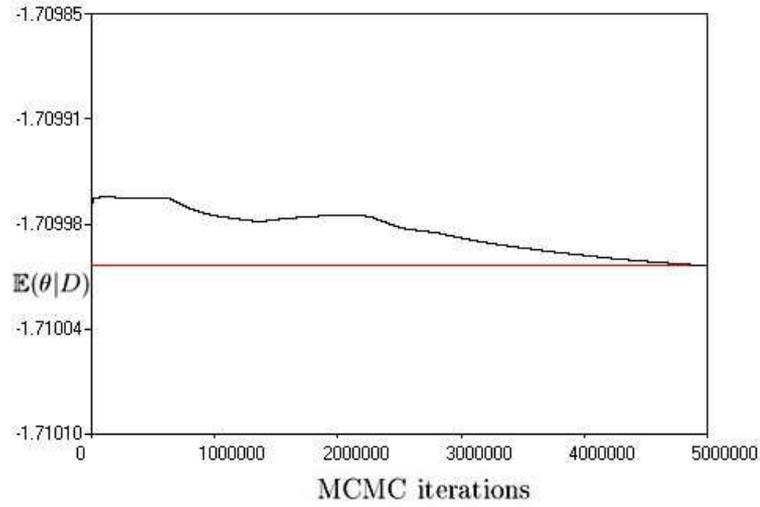}}}\\
\mbox{
\subfigure[]{\includegraphics[angle=0,width=0.6\textwidth]{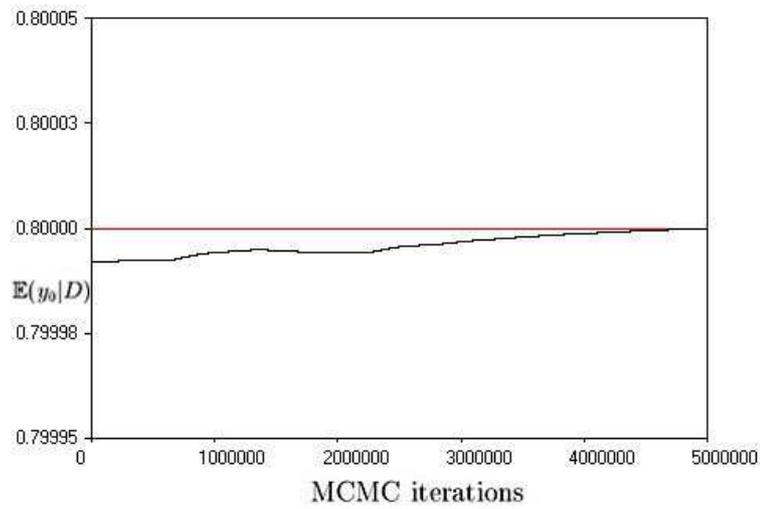}}}\\
\caption{Here the sample size is $\kappa=984$ with $\epsilon=5\times 10^{-5}$ and $\sigma=10^{-8}$.
         The initial conditions for the MCMC scheme in 
         (\ref{ThetaFullConditionals})--(\ref{yFullConditionals}) is the vector 
         $(\theta^*, \tilde{y}_0, \tilde{y}_1, \ldots, \tilde{y}_{\kappa})$ with $\tilde{y}_i$'s coming
         from  (\ref{InverseMapSample}).
         The approximations of the $\theta$ and $y_0$ posterior means
         -- shown in Fig. $5$(a) and $5$(b) respectively -- are converging to
         $\theta=-1.70999996$ and $y_0=0.79999995$ and are correct to $7$ decimal places.}
\label{fig:predictives}
\end{center}
\end{figure}

\newpage

%--------------------------------------------------------------------------------------------------
% TABLE 1
%--------------------------------------------------------------------------------------------------

\begin{table}[b]
\begin{center}
\label{tab:table1}
\begin{tabular}{|c|c|c|} \hline
$\Theta$--interval & $\vartheta$--approximation  & $y_0$--approximation \\ \hline
 (-1.77000, -1.68000)  &   -1.72500  &   0.80256    \\
 (-1.72091, -1.70455)  &   -1.71273  &   0.80073    \\
 (-1.70859, -1.71132)  &   -1.70995  &   0.80026    \\ \hline
\end{tabular}
\end{center}
\caption{The approximations to the true values $\vartheta=-1.71$ and $y_0=0.8$ using the strength 
         maximizing MCMC and a simulated censored data set $D'$ of the form (\ref{QuadraticDataSet}) 
         of size $m=600$.}
\end{table}

%--------------------------------------------------------------------------------------------------
% TABLE 2
%--------------------------------------------------------------------------------------------------

\begin{table}[b]
\begin{center}
\label{tab:table2}
\begin{tabular}{|c|c|c|c|c|c|} \hline
$i$ &    $y_i^*$  & $\bar{y}_i$ & $\tilde{y}_i$ & $|y_i-\bar{y}_i|$ & $|y_i-\tilde{y}_i|$\\ \hline
0   & 0.80000000& 0.80023055  & 0.79999129    & 0.00023055        & 0.00000871\\
1   &-0.09440000&-0.09502181  &-0.09434708    & 0.00062181        & 0.00005292\\
2   & 0.98476157& 0.98446928  & 0.98477906    & 0.00029230        & 0.00001749\\ 
3   &-0.65828166&-0.65722148  &-0.65829647    & 0.00106018        & 0.00001481\\ 
4   & 0.25899758& 0.26034877  & 0.25898394    & 0.00135119        & 0.00001364\\ \hline
\end{tabular}
\end{center}
\caption{The true values $y_i^*=g^{(i)}(\vartheta,y_0^*)$ for $0\le i\le 4$, $\vartheta=-1.71$ and $y_0=0.8$ 
         are on the $2$nd column. The approximations to the candidate values $\bar{y}_i$
         are on the $3$rd column. After the application of $g^{-1}|_D$ using 
         (\ref{InverseMapSample}) with $\kappa=984$, we obtain $\tilde{y}_i$ on the $4$th column. 
         The absolute errors of the 
         previous two approximations are in the last two columns.}
\end{table}

\end{document}